\documentclass[superscriptaddress,aps,twocolumn,longbibliography,nofootinbib]{revtex4-2}
\usepackage{amsmath,amssymb}
\usepackage{mathtools}
\usepackage{graphicx}
\usepackage{dcolumn}
\usepackage{tcolorbox}
\usepackage{enumitem}
\usepackage{float}
\usepackage{bm,dsfont}
\usepackage{multirow}
\usepackage{simpler-wick}
\usepackage{cuted}
\usepackage{color}
\usepackage{lipsum}
\usepackage{ulem}
\usepackage{hyperref}
\hypersetup{
colorlinks = true,
linkcolor = [rgb]{0.70,0.13,0.13},
citecolor = [rgb]{0.13,0.55,0.13},
urlcolor = [rgb]{0.25, 0.41, 0.88}}
\newcommand{\bra}[1]{{\langle #1|}}
\newcommand{\ket}[1]{{|#1 \rangle}}

\newcommand{\ii}{\mathrm{i}}
\newcommand{\id}{\mathds{1}}

\newcommand{\dsE}{\mathbb{E}}

\newcommand{\dsC}{\mathbb{C}}

\newcommand{\Cl}{{\mathcal{C}\ell}}

\newcommand{\scE}{\mathcal{E}}

\newcommand{\scM}{\mathcal{M}}
\newcommand{\scN}{\mathcal{N}}

\newcommand{\scO}{\mathcal{O}}

\newcommand{\scS}{\mathcal{S}}

\newcommand{\scU}{\mathcal{U}}

\newtheorem{theorem}{Theorem}
\newcommand{\Tr}{\operatorname{Tr}}

\newcommand{\var}{\operatorname{Var}}
\newcommand{\cov}{\operatorname{Cov}}
\newcommand{\avg}{\operatorname{avg}}
\renewcommand{\Pr}{\operatorname{Pr}}
\newcommand{\E}{\mathop{\dsE}}

\newcommand{\dia}[3]{\raisebox{#3pt}{\includegraphics[height=#2pt]{dia_#1}}}
\newcommand{\eq}[1]{\begin{equation}#1\end{equation}}
\newcommand{\eqs}[1]{\begin{equation}\begin{split}#1\end{split}\end{equation}}
\newcommand{\eqnref}[1]{Eq.\,\eqref{#1}}
\newcommand{\figref}[1]{Fig.\,\ref{#1}}

\newcommand{\secref}[1]{Sec.\,\ref{#1}}
\newcommand{\appref}[1]{Appendix\,\ref{#1}}
\newcommand{\refcite}[1]{Ref.\,\cite{#1}}
\usepackage[mathlines]{lineno}

\begin{document}


\title{Logical shadow tomography: Efficient estimation of error-mitigated observables}

\author{Hong-Ye Hu}
\thanks{Equal contributions.}
\affiliation{Department of Physics, University of California San Diego, La Jolla, CA 92093, USA}
\affiliation{Department of Physics, Harvard University, 17 Oxford Street, Cambridge, MA 02138, USA}
\affiliation{Quantum Artificial Intelligence Laboratory (QuAIL),
NASA Ames Research Center, Moffett Field, CA, 94035, USA}
\affiliation{USRA Research Institute for Advanced Computer Science (RIACS), Mountain View, CA, 94043, USA}

\author{Ryan LaRose}
\thanks{Equal contributions.}
\affiliation{Department of Computational Mathematics, Science, and Engineering, Michigan State University, East Lansing, MI, 48823, USA}
\affiliation{Quantum Artificial Intelligence Laboratory (QuAIL),
NASA Ames Research Center, Moffett Field, CA, 94035, USA}

\author{Yi-Zhuang You}
\affiliation{Department of Physics, University of California San Diego, La Jolla, CA 92093, USA}

\author{Eleanor Rieffel}
\affiliation{Quantum Artificial Intelligence Laboratory (QuAIL),
NASA Ames Research Center, Moffett Field, CA, 94035, USA}

\author{Zhihui Wang}
\thanks{Corresponding author: \href{mailto:zhihui.wang@nasa.gov}{zhihui.wang@nasa.gov}.}
\affiliation{Quantum Artificial Intelligence Laboratory (QuAIL),
NASA Ames Research Center, Moffett Field, CA, 94035, USA}
\affiliation{USRA Research Institute for Advanced Computer Science (RIACS), Mountain View, CA, 94043, USA}


\begin{abstract}
We introduce a technique to estimate error-mitigated expectation values on noisy quantum computers. Our technique performs shadow tomography on a logical state to produce a memory-efficient classical reconstruction of the noisy density matrix. Using efficient classical post-processing, one can mitigate errors by projecting a general nonlinear function of the noisy density matrix into the codespace. The subspace expansion and virtual distillation can be viewed as special cases of the new framekwork. We show our method is favorable in the quantum and classical resources overhead. Relative to subspace expansion which requires $O\left(2^{N} \right)$ samples to estimate a logical Pauli observable with $[[N, k]]$ error correction code, our technique requires only $O\left(4^{k} \right)$ samples. Relative to virtual distillation, our technique can compute powers of the density matrix without additional copies of quantum states or quantum memory. We present numerical evidence using logical states encoded with up to sixty physical qubits and show fast convergence to error-free expectation values with only $10^5$ samples under 1\% depolarizing noise.

\end{abstract}

\maketitle


\section{Introduction}

Recent advances in experimental quantum information processing have enabled complex simulations of interesting physical systems~\cite{arute_quantum_2020,arute_observation_2020,mi_information_2021,blok_quantum_2021,Ebadi,doi:10.1126/science.abi8794}. A common and crucial element of each experiment is mitigating the effect of errors, a task which is generally referred to as quantum error mitigation (QEM)~\cite{takagi_fundamental_2021}.

A relatively large number of QEM techniques have been proposed in recent literature~\cite{li_efficient_2017,temme_error_2017,mcclean_decoding_2019,czarnik_error_2021,huggins_virtual_2021,Tgate2021,PhysRevResearch.3.033098,cliffordgatemitigation}. Some authors have defined common frameworks which encapsulate one or more of these techniques~\cite{yoshioka_generalized_2021,lowe_unified_2020}. At its core, any QEM technique uses additional quantum resources (qubits, gates, and/or samples) in a clever way to approximate what would happen in an ideal device.

In this work, we are interested in error mitigation in the region between noisy-intermediate scale quantum (NISQ) and fault-tolerant quantum computation. Especially, we present a conceptually simple and practical QEM technique. In our setup, we use quantum error correction code to distribute logical qubits information with multiple physical qubits. Without active quantum error correction which requires extra ancilla qubits and parity check measurements, the only step which happens on a quantum computer is repeatedly sampling from an encoded state to perform shadow tomography~\cite{aaronson_shadow_2018,huang_predicting_2020}. After this, one can use classical post-processing to project out errors as in subspace expansion~\cite{mcclean_decoding_2019} and calculate powers of the density matrix as in virtual distillation~\cite{huggins_virtual_2021}. In addition to enabling both of these methods, our technique requires significantly fewer resources. Especially, we show 1) the quantum gate overhead is independent of the number of logical qubits and we do not require multiple copies of the physical systems, 2) the sample complexity for estimating error mitigated Pauli observables only scales with the number of logical qubits instead of the total number of physical qubits, and 3) there exists an efficient classical algorithm that can post-process data with polynomial time. We show the new method is practical for relatively large systems with numerical evidence.

\section{Logical shadow tomography} \label{sec:logical-shadow-tomography}

\subsection{Motivation}

\begin{figure}[htbp]
    \centering
    \includegraphics[width=0.95\linewidth]{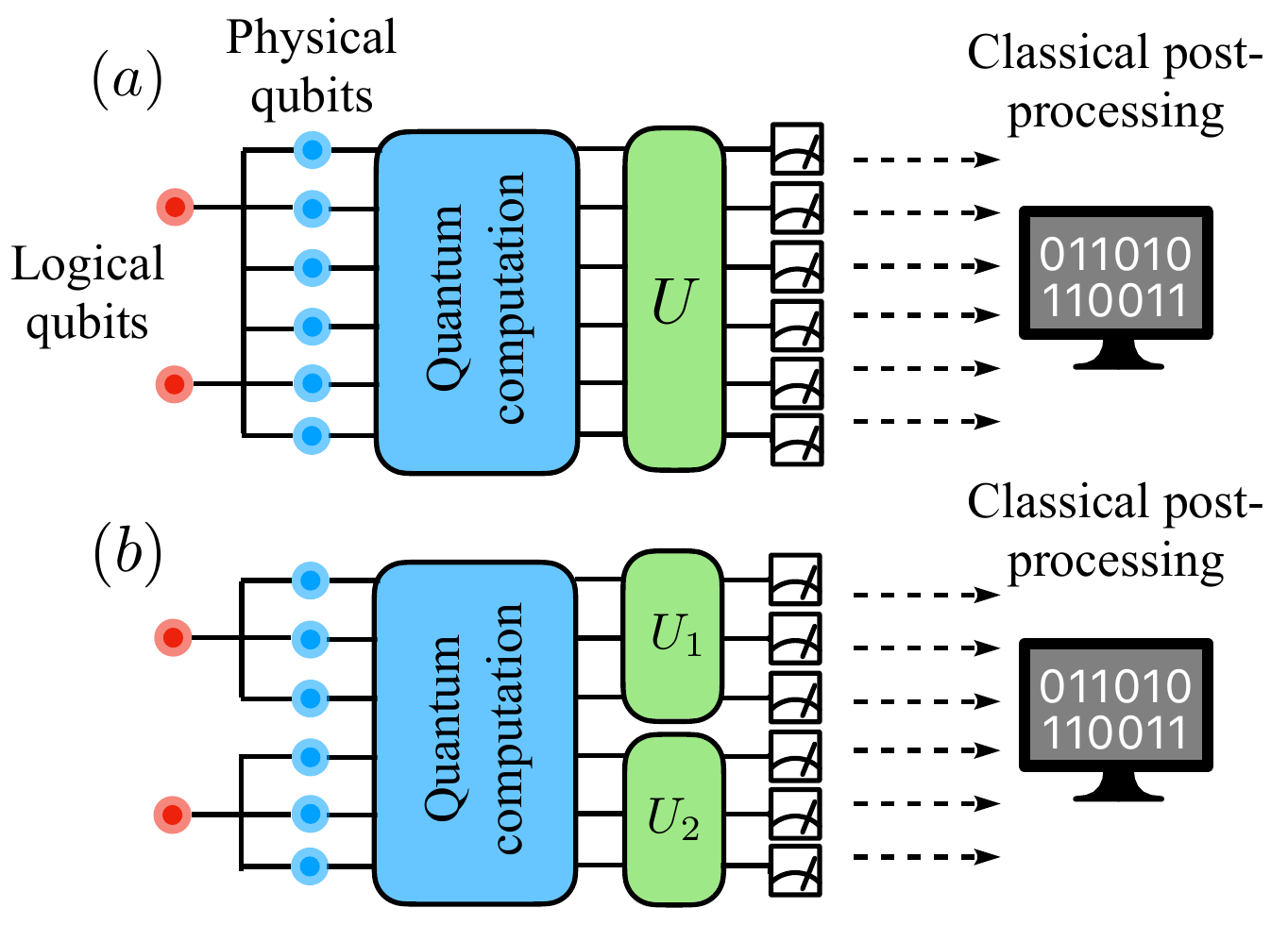}
    \caption{Graphic illustration of logical shadow tomography. (a) Red dots are logical qubits, and blue dots are physical qubits. Logical information is distributed to physical qubits by error correction code, then followed by noisy quantum computation on physical qubits. To get estimation of error mitigated observables, we perform classical shadow tomography on the noisy physical state. Particularly, we can apply random Clifford gates denoted as green blocks from some unitary ensemble $\scU$, and take computational basis measurements. (b) A special case using $[[n,1]]$ code for each logical qubit. In shadow tomography, we apply random unitary from tensor product of Clifford groups $\Cl(2^n)^{\otimes k}$. Additional gate depth will not scale with number of logical qubits $k$, and sample complexity for estimating error mitigated logical Pauli observables is the same as using global Clifford group $\Cl(2^{nk})$.}
    \label{fig:lst_overview}
\end{figure}

One central task in quantum information processing is to estimate the expectation value of an observable $O$ with respect to a pure state $\rho=\ket{\psi}\bra{\psi}$, i.e.~$\langle O\rangle =\bra{\psi}O\ket{\psi}=\Tr(\rho O)$.
If the state is prepared by a quantum device, it can be noisy, and we instead evaluate $\langle O\rangle_\text{noisy}=\Tr(\rho_\scE O)$ by direct measurement, where $\rho_\scE=\scE(\rho)$ denotes the noisy state that is corrupted from the target state $\rho$ by an unknown noisy quantum channel $\scE$. 
Given the corrupted state $\rho_\scE$, the goal of quantum error mitigation is to estimate a quantity $\langle O \rangle_{\text{QEM}}$ that is closer to the target value $\langle O\rangle$ compared to the direct measurement, such that
\begin{equation}
    \left| \langle O \rangle_{\text{QEM}} - \langle O \rangle \right| < \left| \langle O \rangle_{\text{noisy}} - \langle O \rangle \right| .
\end{equation}
QEM can be performed on either physical devices or during data post-processing. 
In this work, we are interested in error mitigation in the region between noisy intermediate-scale quantum (NISQ) and fault-tolerant quantum computation, where the number of qubits is more than a few but cannot fulfill the requirement of fault-tolerance. We ask the question whether we can use those resources cleverly to achieve a more reliable quantum computation. QEM methods related to this idea are subspace expansion~\cite{mcclean_decoding_2019} and the virtual distillation~\cite{huggins_virtual_2021}. The key contribution of this work is to propose the application of classical shadow tomography~\cite{aaronson_shadow_2018,huang_predicting_2020} to reduce the quantum and classical resources needed to perform these QEM schemes, and provide rigorous analysis on its error mitigation capability. 


The subspace expansion approach starts by encoding $k$ logical qubits with $N$ physical qubits via an error correction code $[[N,k]]$. Consider using a stabilizer code defined by a stabilizer group $\scS=\langle S_1,S_2,\cdots,S_{N-k}\rangle$, the code subspace is specified by the projection operator
\eq{\label{eq:Pi}
\Pi=\prod_{j=1}^{N-k}\frac{\id+S_j}{2}=\frac{1}{2^{N-k}}\sum_{M\in\scS}M,}
where $M$ denote group elements in $\scS$ as products of the stabilizers. Errors may occur to the physical state during the quantum information processing, which generally takes the physical state away from the code subspace. Suppose the goal is to estimate the logical observable $O$ only, then even without active error correction, a simple projection of the the corrupted physical state $\rho_\scE$ back to the code subspace can already mitigate the error for the logical observable~\cite{mcclean_decoding_2019}
\begin{equation}\label{eq:SSE}
\langle O\rangle_\text{QEM} = \frac{\Tr(\Pi\rho_\scE \Pi^\dagger O)}{\Tr(\Pi\rho_\scE\Pi^\dagger)}
=\frac{1}{2^{n-k} c}\sum_{M\in\scS}\Tr(\rho_\scE M O),
\end{equation}
where $c=\Tr(\Pi\rho_\scE\Pi^\dagger)$. This amounts to measuring $\langle MO\rangle$ on the corrupted physical state for all elements $M$ in the stabilizer group $\scS$ (or for a majority of $M$ sampled from $\scS$). This approach can quickly become exponentially expensive when $N-k$ becomes large. 

Another approach for QEM goes under the name of virtual distillation. Assuming the target state is a pure state as the leading eigen state of $\rho_\scE$. The sub-leading eigen states of $\rho_\scE$ (as errors orthogonal to the target state) can be suppressed by powering the density matrix $\rho_\scE^m$, and we estimate $\langle O\rangle_\text{QEM}=\Tr(\rho_\scE^m O)/\Tr(\rho_\scE^m)$. Or more generally, a polynomial function $f(\rho_\scE)=c_0\id + c_1\rho_\scE+c_2 \rho_\scE^2+\cdots+c_m \rho_\scE^m$ of $\rho_\scE$ can be considered, with an optimal choice of the coefficients $c_0,\cdots,c_m$ to best mitigate the error, such that
\begin{equation}\label{eq:VD}
\langle O\rangle_\text{QEM}=\frac{\Tr(f(\rho_\scE) O)}{\Tr(f(\rho_\scE))}.
\end{equation}
However, powering a density matrix on quantum devices typically involves making multiple copies of the quantum system, which can be challenging and expensive in quantum resources.

The key observation of this work is that both \eqnref{eq:SSE} and \eqnref{eq:VD} (or their combination) can be efficiently evaluated in the classical post-processing phase after performing the classical shadow tomography~\cite{huang_predicting_2020} on the noisy physical state $\rho_\scE$. The classical shadow tomography uses randomized measurements to extract information from an unknown quantum state, and predicts physical properties about the state efficiently by post-processing collected measurement outcomes on a classical computer. When the measurement basis are chosen to be Pauli or Clifford basis, the post-processing can be made efficient. The code subspace projection and the powering of density matrix can all be implemented efficiently in the post-processing phase by classical computation, given the Clifford nature of the classical shadows. In this way, we can implement the existing error mitigation schemes with significantly reduced quantum and classical resources.

\subsection{Procedure} \label{subsec:lst-procedure}

Let us now introduce our technique which we will refer to as logical shadow tomography (LST). LST consists of the following steps\footnote{Steps 2 and 3 can be aptly summarized by ``perform shadow tomography~\cite{huang_predicting_2020} on the logical state,'' whence \textit{logical shadow tomography}. We explain these steps in detail for completeness.} (see Fig.~\ref{fig:lst_overview} for a graphical overview):

\begin{enumerate}

\item Given a $k$-qubit logical state, encode it into a $N$-qubit physical state by a $[[N, k]]$ stabilizer code. 

\item Perform the quantum information processing on the physical state, the resulting physical state is denoted as $\rho_\scE$. Due to the error accumulated in the processing, $\rho_\scE=\scE(\rho)$ may be corrupted from the ideal result $\rho=\ket{\psi}\bra{\psi}$ by some noisy quantum channel $\scE$. The goal is the mitigate the error for predicting logical observables based on the noisy physical state $\rho_\scE$.

\item Perform classical shadow tomography on the physical state.

\begin{enumerate}[label*=\arabic*.]
\item Apply a randomly sampled unitary $U$ from a unitary ensemble $\scU$ to the physical qubits. Measure all physical qubits in the computational basis to obtain a bit-string $b \in \{0, 1\}^n$. Store $U$, $b$.
    
\item Repeat step 3.1 for $N$ times to obtain a data ensemble $\{(U_s,b_s)\}_{s=1}^{N}$ ($s$ labels the samples in the ensemble).
\end{enumerate}

\item Post-processing the data on a classical computer.

\begin{enumerate}[label*=\arabic*.]
\item Construct the classical shadow ensemble
\begin{equation}\label{eq:shadow}
\Sigma(\rho_\scE)=\{\hat{\rho}_s=\scM^{-1}(U_s^\dagger\ket{b_s}\bra{b_s}U_s)\}_{s=1}^{N},
\end{equation}
where $\scM^{-1}$ is the classical shadow reconstruction map that depends on the unitary ensemble $\scU$.

\item Given any logical observable $O$ (i.e.~$[\Pi,O]=0$) estimate the error-mitigated expectation value by
\eq{\label{eq:LST}
\langle O\rangle_\text{LST}=\frac{\Tr(\Pi f(\rho_\scE)\Pi^\dagger O)}{\Tr(\Pi f(\rho_\scE)\Pi^\dagger )},}
where $\Pi$ is the code subspace projection operator defined in \eqnref{eq:Pi}, and $f(x)=\sum_{p=1}^{m} c_p x^p$ can be a generic polynomial function up to the power $m$. The proposed QEM estimator $\langle O\rangle_\text{LST}$ in \eqnref{eq:LST} combines the subspace expansion \eqnref{eq:SSE} and the virtual distillation \eqnref{eq:VD} approaches. In particular, the numerator $\Tr(\Pi f(\rho_\scE)\Pi^\dagger O)$ is evaluated by
\eq{\label{eq:numerator}
\sum_{p=1}^{m}c_p\E_{\{\hat{\rho}_s\}\in\Sigma(\rho_\scE)^{\times p}}\Tr\bigg(\Pi \Big(\prod_{s=1}^{p}\hat{\rho}_s\Big) \Pi^\dagger O\bigg),}
and the denominator $\Tr(\Pi f(\rho_\scE)\Pi^\dagger)$ is evaluated independently in a similar manner (by replacing $O$ with $\id$).
\end{enumerate}
\end{enumerate}

The map $\mathcal{M}^{-1}$ depends on the unitary ensemble $\mathcal{U}$ for which there are several proposals, e.g. Pauli ensembles, random Clifford circuits, and chaotic dynamics \cite{huang_predicting_2020,PhysRevResearch.4.013054,hu_classical_2021,Ohliger_2013}. In this work, we find the sample complexity for predicting logical Pauli observable is the same between using a full Clifford ensemble $\Cl(2^N)$ as shown in \figref{fig:lst_overview}(a) and a tensor product of Clifford ensemble $\Cl(2^{N/k})^{\otimes k}$ as shown in \figref{fig:lst_overview}(b). In the following, we will focus on the scheme where each qubit is encoded with a $[[n,1]]$ error correction code and apply random unitaries from Clifford group $\Cl(2^n)$ at each logical qubit sector. And the total number of physical qubits is $N=n k$.


\subsection{Analysis\label{subsec:analysis}}

In the previous section, we have outlined the procedure of logical shadow tomography (LST). Here, we will analyze its performance from three perspectives: 1) error mitigation capacity, 2) quantum resources, and 3) classical resources. Particularly, in \emph{error mitigation capacity} subsection, we show how error is suppressed with the code distance $d$ of the error correction code and powers of density matrix. In \emph{quantum resources} subsection, we show the gate overhead is similar to the original proposal of subspace expansion, except for a shallow depth Clifford circuit whose depth does not depend on the number of logical qubits. Compared to virtual distillation, our method only requires one copy of the physical system. In addition, we also show the sample complexity has an exponential reduction compared to the direct implementation of subspace expansion for estimating logical Pauli observables. In \emph{classical resources} subsection, we outline the general classical algorithm for post-processing the data. Particularly, we show there exists fast algorithm for LST with $f(\rho_{\scE})=\rho_{\scE}$ and its algorithm time complexity is $O(N^3)$. This allows our method scale to large system size. 

\subsubsection{Error mitigation capability} \label{sec:error-mitigation-capability}


$\bullet$ {\bf Code space projection}. Let $\mathcal{S}$ be the stabilizer code used in LST. For any correctable error $E$, there exists a stabilizer generator $S$ such that $SE = - ES$ and so
\begin{equation}
    \Pi E|\psi\rangle \propto (\id + S) E |\psi\rangle = E(|\psi\rangle - |\psi\rangle) = 0.\label{eq:local_error}
\end{equation}
Analogously, if no error has occurred then $|\psi\rangle$ is a codeword and
\begin{equation}
    \Pi |\psi\rangle = |\psi\rangle .
\end{equation}
Thus the projector $\Pi$ discards results in which correctable errors have occurred~\cite{mcclean_decoding_2019}. The set of correctable errors is determined by the chosen code. Assume a simple noise model where each qubit is subjected to depolarizing noise with rate $p$. If a single error happened on one qubit. Then it can be projected out given the fact that single Pauli operator is not the stabilizer group. Those errors are non-logical errors.

The code space projection fails when more local error happens and they form a logical operator. The probability of having this failure is $p^d$, where $d$ is the code distance of the error correction code. Mathematically, we can write down
\eqs{
\rho_{\scE}=(1-p)^{N}\rho_0+p \rho_1 + p^2 \rho_2+\dots,\label{eq:new_spectral_decom}
}
where $\rho_0=\ket{\psi_0}\bra{\psi_0}$ is the ideal quantum state, and $N$ is the system size. $\rho_1$ is the quantum state subjected to error happened to one qubit, i.e. \eqs{\rho_1=&(XI\cdot\cdot I)\rho_0(XI\cdot\cdot I)+(IX\cdot\cdot I)\rho_0(IX\cdot\cdot I)+\dots} Similarly, $\rho_2$ is the quantum state subjected to two error happened, i.e. \eqs{\rho_2=&(XIY\cdot\cdot I)\rho_0(XIY\cdot\cdot I)\\& +(IXX\cdot\cdot I)\rho_0(IXX\cdot\cdot I)+\dots} Therefore, for any logical observable $O$, we have 
\eqs{
&\dfrac{\Tr(\Pi\rho_{\scE}\Pi O)}{\Tr(\Pi\rho_{\scE}\Pi)}\\ &=\bra{\psi_0}O\ket{\psi_0}\left[1+O\left(p^d\dfrac{\Tr(\rho_d O)}{\bra{\psi_0}O\ket{\psi_0}}\right)\right].
}
More details can be found in \appref{ap:proof_decom}.

$\bullet$ {\bf Virtual distillation}. For completeness, we will review the virtual distillation theory here. With the spectral decomposition of the noisy density matrix, one can write
\begin{equation}
    \rho_\scE = p_0 |\psi_0\rangle \langle \psi_0| + p_1 |\psi_1\rangle \langle \psi_1 | + \cdots + p_n |\psi_n \rangle \langle \psi_n |,
\end{equation}
where $p_0 > \cdots > p_n$ and $\langle \psi_i | \psi_j \rangle = \delta_{ij}$. For shallow circuit, it is reasonable to assume $|\psi_0\rangle \equiv |\psi\rangle$ is the noiseless state. 

In this case, for any observable $O$ and positive integer $m$ we have~\cite{huggins_virtual_2021,yoshioka_generalized_2021}
\begin{equation}
\begin{aligned}
    \frac{\Tr(\rho_\scE^m O)}{\Tr (\rho_\scE^m )}
        &= \frac{p_0^m\langle \psi_0 | O | \psi_0 \rangle + \sum_i p_i^m \langle \psi_i | O | \psi_i \rangle}{p_0^m + \sum_i p_i^m} \\
        &= \langle \psi | O | \psi \rangle \left[ 1 + O\left( \left(\dfrac{p_1}{p_0}\right)^m\dfrac{\bra{\psi_1}O\ket{\psi_1}}{\bra{\psi}O\ket{\psi}}\right) \right].
\end{aligned}
\end{equation}
Thus, computing the expectation value with the $m$th power of the state suppresses errors exponentially in $m$, a phenomenon which can also be interpreted as artificially cooling the system~\cite{cotler_quantum_2019}. Here we allow for an arbitrary function $f$ acting on the noisy state via its Taylor expansion $f(x)=\sum_{p=1}^{m} c_p x^p$. Including sums of powers up to $m$ (instead of just the highest power $m$) was shown to improve results of numerical experiments in~\cite{yoshioka_generalized_2021}.

$\bullet$ {\bf LST (Combined approach)}. LST has the error mitigation capability of code space projection and virtual distillation. When we combine code space projection with virtual distillation, we expect both code distance $d$ of the error correction code and $m$th power of the density matrix will suppress the error. Especially, when $m=2$ in $\Tr(\Pi \rho_{\epsilon}^m\Pi O)$, i.e. one projects the squared density matrix to code space, the order of error suppression is improved from $\scO(p^{d})$ to $\scO(p^{2d})$. In general, higher order of the power $m$ will lead to stronger error mitigation effect. (See \appref{ap:proof_decom} for more details.)

\subsubsection{Quantum resources\label{subsec:quantum_resources}}

$\bullet$ {\bf Gate overhead}. LST requires additional qubits and gates to encode the logical state, the exact number of which depends on the chosen code. Restricting to stabilizer codes, the logical state preparation only requires implementing Clifford gates, which are presumably easier to implement on NISQ devices compared to universal quantum computation gate set. This encoding overhead is the same as the subspace expansion method. To perform the classical shadow tomography, an element from a unitary ensemble $\mathcal{U}$ is appended to the quantum circuit before measuring all qubits in the computational basis. If $\mathcal{U}$ is chosen to be a global Clifford ensemble, the added circuit depth is $\scO(N)$ with local random unitary gates~\cite{brandao_local_2016}. The gate overhead associated with this can be significant. If $\mathcal{U}$ is chosen to be a tensor-product Pauli ensemble, the added circuit depth is $\scO(1)$~\cite{huang_predicting_2020}. The gate overhead is more affordable. However, the Pauli measurement will increase the sample complexity exponentially for non-local observables. Facing this dilemma between Clifford v.s. Pauli measurement, a recent work \cite{hu_classical_2021} by part of the authors has developed an efficient classical shadow tomography approach for \emph{finite-depth local} Clifford circuits, which can smoothly interpolate between the global Clifford and the local Pauli limit. Using shallow (finite-depth) Clifford circuits for shadow tomography, the gate overhead can be significantly reduced with only a mild increase of sample complexity. \cite{hu_classical_2021} This approach can be combined with our error mitigation technique seamlessly to achieve an optimal balance between gate overhead and sample complexity. For this work, we will use a tensor-product Clifford group $\Cl(2^n)^{\otimes k}$ as shown in \figref{fig:lst_overview} (b). The added circuit depth is $\scO(n)$, where $n$ is the physical qubits in $[[n,1]]$ code for one logical qubit. It is important to notice that the depth of additional circuit is independent of the number of logical qubit $k$.

$\bullet$ {\bf Sample complexity}. 
We now consider number of measurements needed for classical post-processing in LST.
If one uses a $[[n,1]]$ code for each logical qubit and a tensor-product Clifford group $\Cl(2^n)^{\otimes k}$, where $k$ is number of logical qubits, we have the following theorem which dictates the sample complexity of LST.

\begin{theorem}[Details in \appref{ap:proof}]
Let $\rho$ be a $k$ logical qubits quantum state, where each logical qubit is encoded with a $[[n,1]]$ code, and $\Pi$ be the associated projection operator to the code subspace. Then one needs $\scO(\log(M)4^{k}/\epsilon\delta^{2})$ samples to produce an estimation $\tilde{O}_i$ of $\Tr(\Pi \rho\Pi O_{i})$ with $\{i=1,\dots,M\}$ of logical Pauli observables $O_i$ such that
\begin{equation} \label{eqn:shadow-tomography-accuracy}
    \text{Pr} \left(
        \left|\tilde{O}_i - \Tr(\Pi \rho\Pi O_{i}) \right| \ge \epsilon
    \right)
    \leq \delta.
\end{equation}
\end{theorem}
The result~\eqref{eqn:shadow-tomography-accuracy} applies to both the numerator and denominator of the LST estimate \eqnref{eq:LST} separately.  We emphasize that the number of samples does not depend on the total number of physical qubits $nk$, but only depends on the number of logical qubits $k$, and scales as $\scO(4^k)$. Compared to direct implementation of subspace expansion whose sample complexity is $\scO(2^{nk})$, our method dramatically reduces the sample complexity. 

In \refcite{mcclean_decoding_2019}, the authors also mentioned stochastic sampling of the stabilizer group elements to reduce the sample complexity at the price of losing the error mitigation capacity. Our approach bypasses the need to sample the all elements in the stabilizer group, as we can implement the projection operator directly and efficiently by data post-processing. Our advantage will be even more apparent if one uses larger $n$ error correction code $[[n,1]]$, which provides larger code distance and better error mitigation capacity.

\subsubsection{Classical resources}

In this section, we show that there exists efficient algorithm for classical post-processing. Especially, for LST with $f(\rho_\scE)=\rho_\scE$ (no virtual distillation), the classical post-processing can be performed with polynomial classical memory and time.

After sampling, we have the classical shadow \eqnref{eq:shadow} consisting of $N$ stabilizer states $U^{\dagger}\ket{b}\bra{b}U$, the observable $O$, and the projector $\Pi$. We need to estimate the numerator and denominator of \eqnref{eq:LST}, both of which can be written
\begin{equation}
    \Tr [ f(\rho_\scE) \Gamma]
\end{equation}
where $\Gamma = \Pi O$ for the numerator and $\Gamma = \Pi$ for the denominator. Let $m$ be the highest degree of the Taylor expansion of $f$. As in~\cite{huang_predicting_2020}, we can evaluate the expectation of this term via
\begin{equation} \label{eqn:trace-m-power}
    \mathbb{E} \Tr [ \hat{\rho}_{i_1}  \cdots  \hat{\rho}_{i_m} \Gamma],
\end{equation}
where $\{\hat{\rho}_{i_1}\dots\hat{\rho}_{i_m}$ are independent samples. A single classical shadow $\hat{\rho}_i$ requires $\scO(N^2)$ classical memory to store (where $N=nk$ is the total number of qubits), so the argument inside the trace in~\eqref{eqn:trace-m-power} requires $\scO(N^2m)$ classical memory~\cite{aaronson_improved_2004}. In \appref{ap:trace}, we showed that the evaluation of \eqref{eqn:trace-m-power} boils down to evaluating the general form 
\eq{\label{eq:general_Tr}\Tr\bigg(\prod_{j=1}^{l}(a_j\id+b_j M_j)\bigg),} 
where $M_j$ are Pauli operators. This can be solved by finding the null space $\scN_A$ of a binary matrix representation of Pauli operators $\{M_j\}$. By simple counting argument (see \appref{ap:trace} for more details), we see the evaluation of \eqnref{eqn:trace-m-power} has time complexity upper bounded by $\scO\left(mN^2(mN+N-k+1)+m |\scN_A|\right)$, when $m\geq 2$, where $k$ is number of logical qubits, $N=nk$ is the total number of physical qubits, $m$ is power of density matrix, and $|\scN_A|$ is the volume of binary null space $\scN_A$ determined by a set of classical shadows $\{\hat{\rho}_{i_1},\dots,\hat{\rho}_{i_m}\}$. When $m$ is large, the null space $\scN_A$ can be very large. In practice, the evaluation of \eqref{eqn:trace-m-power} can be slow given the time complexity is proportional to $\scO(m|\scN_A|)$ for $m\geq 2$. 

We also find an improved classical algorithm for $m=1$ with time complexity $\scO(N^3)$. When $m=1$, we would like to evaluate $\Tr(\Pi \rho\Pi O)$, where $\rho$ is proportional to a stabilizer state, which can be represented using stabilizer tableau. The intuition behind the efficient algorithm is that a stabilizer state can be efficiently projected by another stabilizer group projector by updating its stabilizer tableau. We leave the detail of the algorithm to \appref{ap:stab_projection}. 

In conclusion, at least for the $m=1$ (no virtual distillation) case, the post-processing complexity is polynomial $\scO(N^3)$ in the total qubit number $N$.

\section{Numerical results}

In Sec.~\ref{subsec:analysis}, we discussed the error mitigation capabilities, quantum resources, and classical resources of LST. One should notice that the discussion of sample complexity is mainly focused on estimation of $\Tr(\Pi\rho\Pi^{\dagger}O)$. While in practice, we would like to estimate the ratio of $\Tr(\Pi\rho\Pi^{\dagger}O)/\Tr(\Pi\rho\Pi^{\dagger})$. The sample complexity of the ratio does not have a closed form in general. This problem is not unique to our approach. It also exists for the subspace expansion \cite{mcclean_decoding_2019} and virtual distillation \cite{huggins_virtual_2021}. 

Nevertheless, in the following, we will use numerical experiments to investigate the sample complexity. We demonstrate the performance of LST through numerical simulation of large systems. In particular, we find LST outperforms the direct implementation of the subspace expansion and the sample complexity scaling is very close to our theoretical prediction in small noise region.

\subsection{Pseudo-threshold with the $[[5, 1, 3]]$ code}

\begin{figure}[htbp]
    \centering
    \includegraphics[width=0.93\columnwidth]{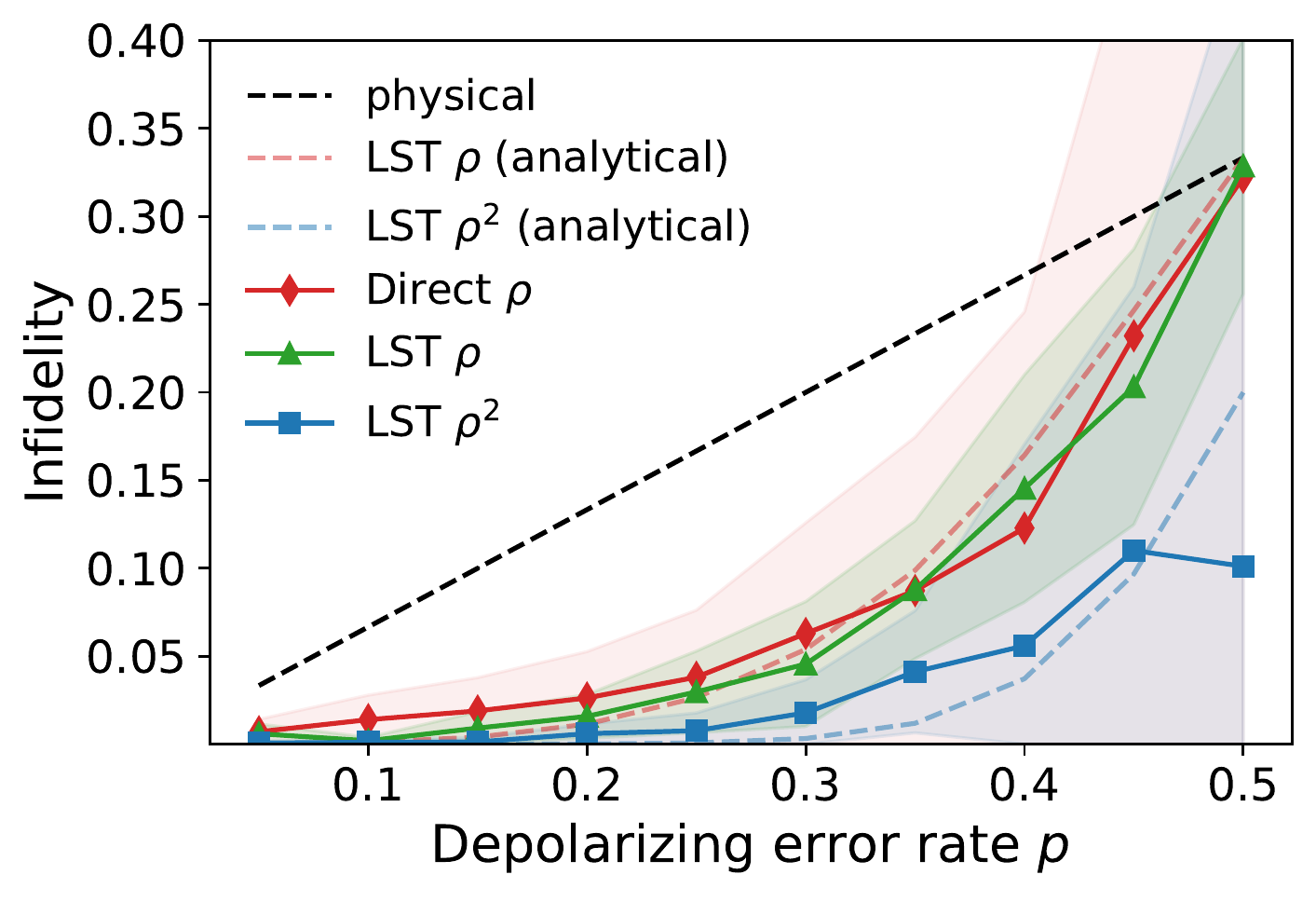}
    \caption{LST with the $[[5, 1, 3]]$ code. Here, $|\psi\rangle$ is taken to be the logical $\ket{\bar{0}}$ and we estimate infidelity $1-F$ with samples. The dashed black line shows the physical infidelity, i.e., the noisy expectation value of single qubit without any encoding. The green and blue dashed line are analytical performance of logical shadow tomography with $f(\rho_{\scE})=\rho_{\scE}$ and $f(\rho_{\scE}^2)=\rho_{\scE}^2$ respectively. The red dots and red shaded area indicates the mean value and standard deviation of error mitigation with $f(\rho_{\scE}) = \rho_{\scE}$ by direct implementation of subspace expansion with 3000 measurements. The green line and green shaded area indicate the mean value and standard deviation with $f(\rho_{\scE}) = \rho_{\scE}$ and 3000 measurements by LST. And the performance of LST with $f(\rho_{\scE}) = \rho_{\scE}^{2}$ is indicated by blue line and blue shaded area.}
    
    \label{fig:lst_513}
\end{figure}

We first consider a simple example with one logical qubit encoded in five physical qubits with the $[[5,1,3]]$ stabilizer code. Each physical qubit is subjected to depolarizing noise with error rate $p$. The same model is considered in the subspace expansion literature \cite{mcclean_decoding_2019}, which shows pseudo-threshold $p=0.5$. Here, we want to compare the practical performance of logical shadow tomography and direct measurement by subspace expansion.

We evaluate \eqnref{eq:LST} with $f(\rho_{\scE}) = \rho_{\scE}$ and $f(\rho_{\scE}) = \rho_{\scE}^2$. The results are shown in Fig.~\ref{fig:lst_513}. Here, the dashed black line shows the infidelity without any error mitigation (the ``physical'' curve), and the dashed colored lines show the infidelity using LST. We see that the LST estimates have lower infidelity than the physical curve, showing that errors are indeed being mitigated. LST with $f(\rho_{\scE}) = \rho_{\scE}^2$ outperforms LST with $f(\rho_{\scE}) = \rho_{\scE}$ showing that the combination of codespace projection and virtually distillation outperforms only projecting into the codespace. This phenomenon agrees with the behavior of the error mitigation capability $p^{md}$ with $m=1,2$ argued in \ref{sec:error-mitigation-capability} (See \appref{ap:proof_decom} for proof details). In addition to expected performance, we also care about sample efficiency, since one major contribution of our work is showing the exponential reduction in sample complexity with LST.

To show this practical advantage, we collect 3000 measurements results and use the data to estimate error mitigated value. The colored lines/points shows the mean values of the estimation and the colored regions shows the standard deviation to the mean value. For $f(\rho_{\scE})=\rho_{\scE}$, LST is the same as the subspace expansion. If we implement the direct subspace expansion by measuring every Pauli observables, each Pauli observable is measured around $100$ times. The red points/line and red shaded region shows the result of direct implementation of subspace expansion. Given each Pauli observable is not measured many times, small error in the denominator can cause large error of the ratio. We see the standard deviation is very huge. The mean estimation and standard deviation of LST is shown as green points/line and green shaded region. In contrast to direct implementation of subspace expansion, LST has a much smaller fluctuation with the same amount of measurements. This shows the practical advantage of our method. In addition, the blue points/line and blue shaded region shows the result of LST with $f(\rho_{\scE})=\rho_{\scE}^2$. With the same amount of data, we see it will suppress the noise even more.

\subsection{Convergence vs. code size}

\begin{figure}[htbp]
    \centering
    \includegraphics[width=0.8\columnwidth]{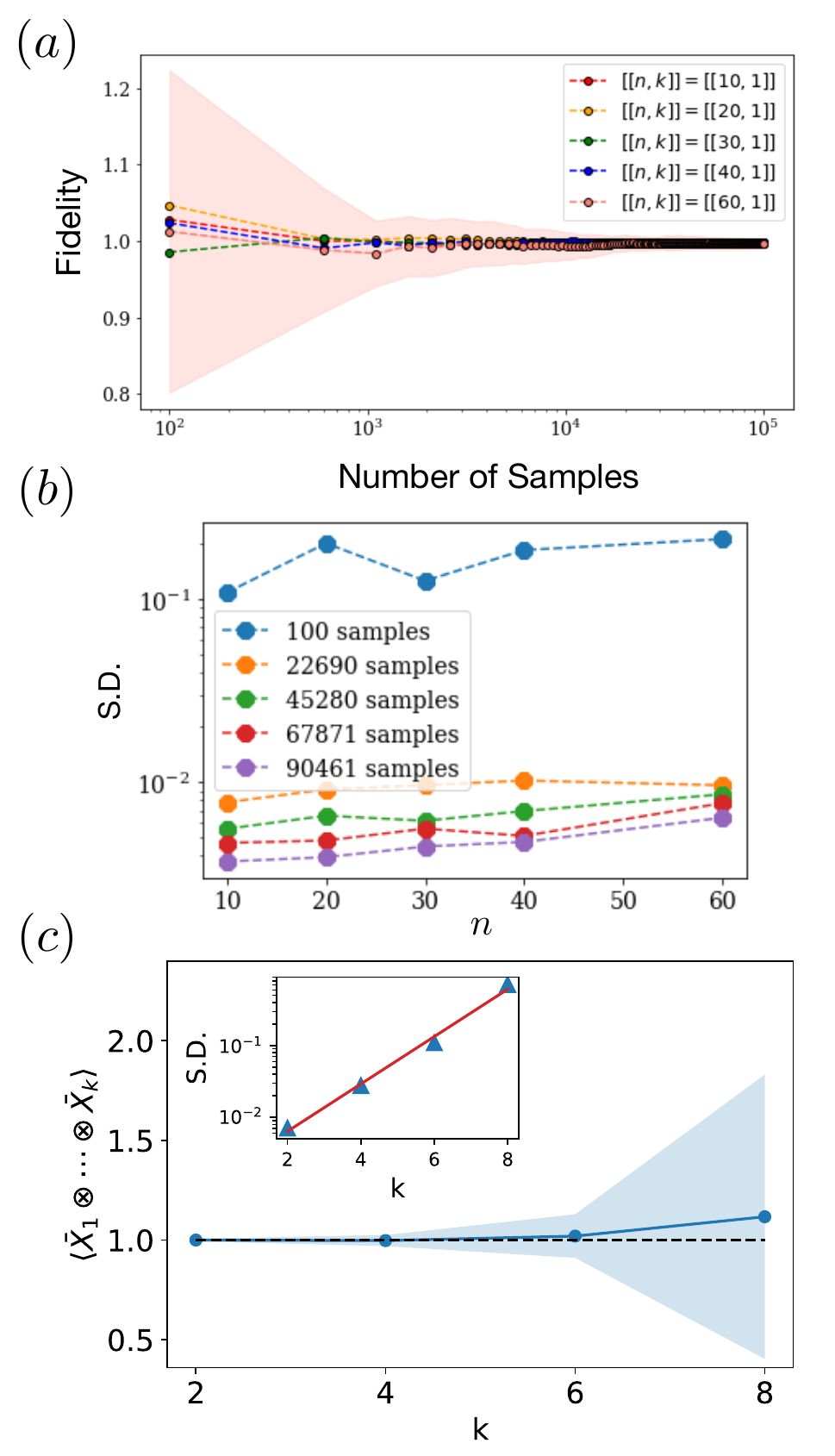}
    \caption{Scaling study of LST with $f(\rho_{\epsilon})=\rho_{\epsilon}$. In all figures, each physical qubit is subjected to 1$\%$ depolarizing noise. (a) LST estimated fidelity vs. number of samples from $10^2$ - $10^5$ with various $[[n,1]]$ code sizes. The noiseless fidelity value of $1.0$ is shown with the dashed black line. For all code sizes up to $n = 60$ physical qubits, the LST estimate converges to the true noiseless value. Codes used are the minimum distance constructions from~\cite{grassl_bounds_nodate}.(b) Standard deviation vs. number of physical qubits $n$. The standard deviation of estimation doesn't scale with number of encoding physical qubits. (c) Mean value and standard deviation scaling vs. number of logical qubits $k$. Each logical qubit is encoded with $[[5,1,3]]$ code, and the state is prepared as logical GHZ state $\ket{\bar{0}\dots\bar{0}}+\ket{\bar{1}\dots \bar{1}}$. We see standard deviation scales exponentially with number of logical qubits $k$ as predicted.}
    \label{fig:lst_convergence}
\end{figure}

As we have pointed out in \secref{subsec:quantum_resources}, the sample complexity for estimating $\Tr(\Pi \rho\Pi^{\dagger} O)$ for logical Pauli observables $O$ only scales with number of logical qubits $k$ as $O(4^k)$ but does not scale with the number of encoding qubits $n$ of the $[[n,1]]$ code for each logical qubit. In practice, we will estimate error mitigated values as a ratio, i.e. $\Tr(\Pi \rho\Pi^{\dagger} O)/\Tr(\Pi \rho\Pi^{\dagger})$. Since there is no closed form for the statistical fluctuation of the above ratio. We will investigate the sample complexity of it via numerical simulation. Interestingly, we find the sample complexity of the ratio agrees well with our theoretical analysis in small noise region.

We now consider $[[n, 1]]$ codes for one logical qubit and vary the number of physical qubits ranging from $n = 10$ to $n = 60$. For quantum noise, each physical qubit is subjected to $1\%$ depolarizing noise in all the following experiments. The LST estimated fidelity (using $f(\rho_{\epsilon}) = \rho_\epsilon$ for all code sizes) vs. number of samples is shown in Fig.~\ref{fig:lst_convergence} (a). Using a relatively small number of samples (at most $10^5$), all LST values converge to the noiseless fidelity. Note that the direct implementation of subspace expansion with the full projection $\Pi$ as used here would require at least $2^{59}$ samples, a number infeasible to implement in any practical experiment. In addition, we estimate the standard deviation of each predicted point by LST using the bootstrap method, and the result is shown in \figref{fig:lst_convergence} (b). We see the standard deviation does not show strong dependence of number of encoding qubits $n$, and it indicates the sample complexity does not increase much if one increase $n$.

In addition, we also study the sample complexity scaling with number of logical qubits $k$. Particularly, we use $[[5,1,3]]$ code to encode each logical qubit and prepare a multi-logical qubits GHZ state, $(\ket{\bar{0}\dots\bar{0}}+\ket{\bar{1}\dots \bar{1}})/\sqrt{2}$. In \figref{fig:lst_convergence} (c), the blue dots/line shows the estimated mean value for logical operator $\bar{X}_1\otimes \cdots\otimes \bar{X}_k$, and the blue shaded area indicates the standard deviation. Especially, in the inset, we see the standard deviation increases exponentially with number of logical qubits $k$. This result also agrees well with our sample complexity analysis, even though the analysis focuses on estimating $\Tr(\Pi \rho\Pi^{\dagger} O)$.

Through the large scale numerical simulation, we find LST indeed outperforms previous methods in terms of sample complexity. Interestingly, we also find the sample complexity agrees well with the theoretical analysis in small noise region. This indicates the sample complexity of LST does not scale much with the number of physical encoding qubits $n$ with $[[n,1]]$ error correction code and only scales with number of logical qubits $k$.

\section{Discussion}

We have presented a procedure for estimating error-mitigated observables on noisy quantum computers. Our procedure is flexible enough to be performed on virtually any quantum computer: the only additional quantum resources needed are qubits and Clifford gates for encoding the logical state. After sampling from the logical state, a classical computer processes the obtained classical shadow to return the error-mitigated expectation value. In the analysis, we show if the error correction code has code distance $d$ and $f(\rho_\scE)=\rho_\scE$ for LST, then error will be suppressed to $\scO(p^{d})$, assuming independent depolarizing noise with rate $p$ on each physical qubit. And we also show higher power of density matrix will further improve the performance. For sample complexity, we show it scales only with number of logical qubits $k$ but not the number of encoding physical qubits $n$. This result is also supported with large scale numerics. In addition, we provide efficient classical algorithms to post-process the classical shadow data and remark that this post-processing can be easily parallelized for practical efficiency in real-world experiments.

With respect to error mitigation, our procedure provides an experimentally simple procedure to carry out proposed error mitigation techniques~\cite{mcclean_decoding_2019, huggins_virtual_2021} at scale. In particular, we have demonstrated subspace expansion with up to $n = 60$ physical qubits encoding a single logical qubit, i.e., a stabilizer group with $2^{59}$ elements, an experiment which would be practically infeasible with the direct or stochastic sampling schemes of~\cite{mcclean_decoding_2019,yoshioka_generalized_2021}. We have also implemented virtual distillation~\cite{huggins_virtual_2021} without expensive swap operations to compute powers of the density matrix. Rather, our procedure uses the same quantum circuit to evaluate the error-mitigated expectation with any function $f(\rho_\scE)$ of the noisy state $\rho_\scE$; the only difference is in classical post-processing (and number of shadows needed). Note that virtual distillation without subspace expansion can be implemented in our protocol by using a trivial code (i.e., not encoding a logical state). Further, beyond making both of these techniques significantly more practical to implement, our procedure enables them to be composed with one another, and we have shown numerically the composition of both techniques results in further reduction of errors. Additional error mitigation techniques which act on the noisy state, e.g., those in~\cite{yoshioka_generalized_2021}, may also be implementable with our framework.

In our analysis, we assumed the Clifford circuit in the classical shadow part is noise-free. If there are noise in the Clifford circuit part, it can be mitigated if the noise is independent of the Clifford gates, as in \cite{robust_shadow2021,2020arXiv201111580E,2022arXiv220203272B}, where similar idea was used for randomized benchmarking~\cite{PhysRevLett.106.180504,PRXQuantum.2.010351}.

Shadow tomography since proposed in~\cite{huang_predicting_2020} has found a number of applications in quantum information processing, including the recently proposed process tomography~\cite{levy_classical_2021}, and avoiding barren plateau in variational quantum circuits~\cite{2022arXiv220108194S}. This work constitutes an application in the error mitigation realm.  We are optimistic our procedure will be effective on current and near-term quantum computers for a variety of experiments on relatively large systems.

In a recent literature \cite{cliffordgatemitigation}, the authors discuss an error mitigation method to implement low noise logical gates for error correction codes. In comparison, our method focuses on the error mitigation of the final quantum states of the logical qubits.

While we were finishing the writing of the paper we learned of related work ``\emph{Shadow Distillation: Quantum Error Mitigation with Classical Shadows for Near-Term Quantum Processors}'' by Alireza Seif, Ze-Pei Cian, Sisi Zhou, Senrui Chen, and
Liang Jiang~\cite{AlirezaQEM}.

\vspace{1em}

\textbf{Acknowledgements} \hspace{1em} We are grateful for support from the NASA Ames Research Center and from the DARPA ONISQ program under interagency agreement IAA 8839, Annex 114.  H.Y.H. is supported by the USRA Feynman Quantum Academy funded by the NAMS R$\&$D Student Program and a UC Hellman Fellowship. R.L. acknowledges support from a NASA Space Technology Graduate Research Fellowship. H.Y.H. and Y.Z.Y. are also supported by a UC Hellman Fellowship. Z.W. is supported by USRA NASA Academic Mission Service (NNA16BD14C).

\bibliographystyle{unsrt}
\bibliography{refs}

\begin{thebibliography}{10}

\bibitem{arute_quantum_2020}
Frank Arute, Kunal Arya, Ryan Babbush, Dave Bacon, Joseph~C. Bardin, Rami
  Barends, Sergio Boixo, Michael Broughton, Bob~B. Buckley, David~A. Buell,
  Brian Burkett, Nicholas Bushnell, Yu~Chen, Zijun Chen, Ben Chiaro, Roberto
  Collins, William Courtney, Sean Demura, Andrew Dunsworth, Edward Farhi,
  Austin Fowler, Brooks Foxen, Craig Gidney, Marissa Giustina, Rob Graff, Steve
  Habegger, Matthew~P. Harrigan, Alan Ho, Sabrina Hong, Trent Huang, L.~B.
  Ioffe, Sergei~V. Isakov, Evan Jeffrey, Zhang Jiang, Cody Jones, Dvir Kafri,
  Kostyantyn Kechedzhi, Julian Kelly, Seon Kim, Paul~V. Klimov, Alexander~N.
  Korotkov, Fedor Kostritsa, David Landhuis, Pavel Laptev, Mike Lindmark,
  Martin Leib, Erik Lucero, Orion Martin, John~M. Martinis, Jarrod~R. McClean,
  Matt McEwen, Anthony Megrant, Xiao Mi, Masoud Mohseni, Wojciech Mruczkiewicz,
  Josh Mutus, Ofer Naaman, Matthew Neeley, Charles Neill, Florian Neukart,
  Hartmut Neven, Murphy~Yuezhen Niu, Thomas~E. O'Brien, Bryan O'Gorman, Eric
  Ostby, Andre Petukhov, Harald Putterman, Chris Quintana, Pedram Roushan,
  Nicholas~C. Rubin, Daniel Sank, Kevin~J. Satzinger, Andrea Skolik, Vadim
  Smelyanskiy, Doug Strain, Michael Streif, Kevin~J. Sung, Marco Szalay, Amit
  Vainsencher, Theodore White, Z.~Jamie Yao, Ping Yeh, Adam Zalcman, and Leo
  Zhou.
\newblock Quantum {Approximate} {Optimization} of {Non}-{Planar} {Graph}
  {Problems} on a {Planar} {Superconducting} {Processor}.
\newblock {\em arXiv:2004.04197 [quant-ph]}, April 2020.
\newblock arXiv: 2004.04197.

\bibitem{arute_observation_2020}
Frank Arute, Kunal Arya, Ryan Babbush, Dave Bacon, Joseph~C. Bardin, Rami
  Barends, Andreas Bengtsson, Sergio Boixo, Michael Broughton, Bob~B. Buckley,
  David~A. Buell, Brian Burkett, Nicholas Bushnell, Yu~Chen, Zijun Chen, Yu-An
  Chen, Ben Chiaro, Roberto Collins, Stephen~J. Cotton, William Courtney, Sean
  Demura, Alan Derk, Andrew Dunsworth, Daniel Eppens, Thomas Eckl, Catherine
  Erickson, Edward Farhi, Austin Fowler, Brooks Foxen, Craig Gidney, Marissa
  Giustina, Rob Graff, Jonathan~A. Gross, Steve Habegger, Matthew~P. Harrigan,
  Alan Ho, Sabrina Hong, Trent Huang, William Huggins, Lev~B. Ioffe, Sergei~V.
  Isakov, Evan Jeffrey, Zhang Jiang, Cody Jones, Dvir Kafri, Kostyantyn
  Kechedzhi, Julian Kelly, Seon Kim, Paul~V. Klimov, Alexander~N. Korotkov,
  Fedor Kostritsa, David Landhuis, Pavel Laptev, Mike Lindmark, Erik Lucero,
  Michael Marthaler, Orion Martin, John~M. Martinis, Anika Marusczyk, Sam
  McArdle, Jarrod~R. McClean, Trevor McCourt, Matt McEwen, Anthony Megrant,
  Carlos Mejuto-Zaera, Xiao Mi, Masoud Mohseni, Wojciech Mruczkiewicz, Josh
  Mutus, Ofer Naaman, Matthew Neeley, Charles Neill, Hartmut Neven, Michael
  Newman, Murphy~Yuezhen Niu, Thomas~E. O'Brien, Eric Ostby, Bálint Pató,
  Andre Petukhov, Harald Putterman, Chris Quintana, Jan-Michael Reiner, Pedram
  Roushan, Nicholas~C. Rubin, Daniel Sank, Kevin~J. Satzinger, Vadim
  Smelyanskiy, Doug Strain, Kevin~J. Sung, Peter Schmitteckert, Marco Szalay,
  Norm~M. Tubman, Amit Vainsencher, Theodore White, Nicolas Vogt, Z.~Jamie Yao,
  Ping Yeh, Adam Zalcman, and Sebastian Zanker.
\newblock Observation of separated dynamics of charge and spin in the
  {Fermi}-{Hubbard} model.
\newblock {\em arXiv:2010.07965 [quant-ph]}, October 2020.
\newblock arXiv: 2010.07965.

\bibitem{mi_information_2021}
Xiao Mi, Pedram Roushan, Chris Quintana, Salvatore Mandra, Jeffrey Marshall,
  Charles Neill, Frank Arute, Kunal Arya, Juan Atalaya, Ryan Babbush, Joseph~C.
  Bardin, Rami Barends, Andreas Bengtsson, Sergio Boixo, Alexandre Bourassa,
  Michael Broughton, Bob~B. Buckley, David~A. Buell, Brian Burkett, Nicholas
  Bushnell, Zijun Chen, Benjamin Chiaro, Roberto Collins, William Courtney,
  Sean Demura, Alan~R. Derk, Andrew Dunsworth, Daniel Eppens, Catherine
  Erickson, Edward Farhi, Austin~G. Fowler, Brooks Foxen, Craig Gidney, Marissa
  Giustina, Jonathan~A. Gross, Matthew~P. Harrigan, Sean~D. Harrington, Jeremy
  Hilton, Alan Ho, Sabrina Hong, Trent Huang, William~J. Huggins, L.~B. Ioffe,
  Sergei~V. Isakov, Evan Jeffrey, Zhang Jiang, Cody Jones, Dvir Kafri, Julian
  Kelly, Seon Kim, Alexei Kitaev, Paul~V. Klimov, Alexander~N. Korotkov, Fedor
  Kostritsa, David Landhuis, Pavel Laptev, Erik Lucero, Orion Martin, Jarrod~R.
  McClean, Trevor McCourt, Matt McEwen, Anthony Megrant, Kevin~C. Miao, Masoud
  Mohseni, Wojciech Mruczkiewicz, Josh Mutus, Ofer Naaman, Matthew Neeley,
  Michael Newman, Murphy~Yuezhen Niu, Thomas~E. O'Brien, Alex Opremcak, Eric
  Ostby, Balint Pato, Andre Petukhov, Nicholas Redd, Nicholas~C. Rubin, Daniel
  Sank, Kevin~J. Satzinger, Vladimir Shvarts, Doug Strain, Marco Szalay,
  Matthew~D. Trevithick, Benjamin Villalonga, Theodore White, Z.~Jamie Yao,
  Ping Yeh, Adam Zalcman, Hartmut Neven, Igor Aleiner, Kostyantyn Kechedzhi,
  Vadim Smelyanskiy, and Yu~Chen.
\newblock Information {Scrambling} in {Computationally} {Complex} {Quantum}
  {Circuits}.
\newblock {\em Science}, page eabg5029, October 2021.
\newblock arXiv: 2101.08870.

\bibitem{blok_quantum_2021}
M.~S. Blok, V.~V. Ramasesh, T.~Schuster, K.~O'Brien, J.~M. Kreikebaum,
  D.~Dahlen, A.~Morvan, B.~Yoshida, N.~Y. Yao, and I.~Siddiqi.
\newblock Quantum {Information} {Scrambling} in a {Superconducting} {Qutrit}
  {Processor}.
\newblock {\em Physical Review X}, 11(2):021010, April 2021.
\newblock arXiv: 2003.03307.

\bibitem{Ebadi}
Sepehr Ebadi, Tout~T. Wang, Harry Levine, Alexander Keesling, Giulia Semeghini,
  Ahmed Omran, Dolev Bluvstein, Rhine Samajdar, Hannes Pichler, Wen~Wei Ho,
  Soonwon Choi, Subir Sachdev, Markus Greiner, Vladan Vuleti{\'c}, and
  Mikhail~D. Lukin.
\newblock Quantum phases of matter on a 256-atom programmable quantum
  simulator.
\newblock {\em Nature}, 595(7866):227--232, 2021.

\bibitem{doi:10.1126/science.abi8794}
G.~Semeghini, H.~Levine, A.~Keesling, S.~Ebadi, T.~T. Wang, D.~Bluvstein,
  R.~Verresen, H.~Pichler, M.~Kalinowski, R.~Samajdar, A.~Omran, S.~Sachdev,
  A.~Vishwanath, M.~Greiner, V.~Vuletić, and M.~D. Lukin.
\newblock Probing topological spin liquids on a programmable quantum simulator.
\newblock {\em Science}, 374(6572):1242--1247, 2021.

\bibitem{takagi_fundamental_2021}
Ryuji Takagi, Suguru Endo, Shintaro Minagawa, and Mile Gu.
\newblock Fundamental limits of quantum error mitigation.
\newblock {\em arXiv:2109.04457 [quant-ph]}, October 2021.
\newblock arXiv: 2109.04457.

\bibitem{li_efficient_2017}
Ying Li and Simon~C. Benjamin.
\newblock Efficient {Variational} {Quantum} {Simulator} {Incorporating}
  {Active} {Error} {Minimization}.
\newblock {\em Physical Review X}, 7(2):021050, June 2017.

\bibitem{temme_error_2017}
Kristan Temme, Sergey Bravyi, and Jay~M. Gambetta.
\newblock Error mitigation for short-depth quantum circuits.
\newblock {\em Physical Review Letters}, 119(18):180509, November 2017.
\newblock arXiv: 1612.02058.

\bibitem{mcclean_decoding_2019}
Jarrod~R. McClean, Zhang Jiang, Nicholas~C. Rubin, Ryan Babbush, and Hartmut
  Neven.
\newblock Decoding quantum errors with subspace expansions.
\newblock {\em arXiv:1903.05786 [physics, physics:quant-ph]}, March 2019.
\newblock arXiv: 1903.05786.

\bibitem{czarnik_error_2021}
Piotr Czarnik, Andrew Arrasmith, Patrick~J. Coles, and Lukasz Cincio.
\newblock Error mitigation with {Clifford} quantum-circuit data.
\newblock {\em arXiv:2005.10189 [quant-ph]}, February 2021.
\newblock arXiv: 2005.10189.

\bibitem{huggins_virtual_2021}
William~J. Huggins, Sam McArdle, Thomas~E. O'Brien, Joonho Lee, Nicholas~C.
  Rubin, Sergio Boixo, K.~Birgitta Whaley, Ryan Babbush, and Jarrod~R. McClean.
\newblock Virtual {Distillation} for {Quantum} {Error} {Mitigation}.
\newblock {\em arXiv:2011.07064 [quant-ph]}, August 2021.
\newblock arXiv: 2011.07064.

\bibitem{Tgate2021}
Christophe Piveteau, David Sutter, Sergey Bravyi, Jay~M. Gambetta, and Kristan
  Temme.
\newblock Error mitigation for universal gates on encoded qubits.
\newblock {\em Physical Review Letters}, 127(20), Nov 2021.

\bibitem{PhysRevResearch.3.033098}
Angus Lowe, Max~Hunter Gordon, Piotr Czarnik, Andrew Arrasmith, Patrick~J.
  Coles, and Lukasz Cincio.
\newblock Unified approach to data-driven quantum error mitigation.
\newblock {\em Phys. Rev. Research}, 3:033098, Jul 2021.

\bibitem{cliffordgatemitigation}
Andrew~Thomas Arrasmith, Piotr~Jan Czarnik, Patrick~Joseph Coles, and Lukasz
  Cincio.
\newblock Error mitigation with clifford quantum-circuit data.
\newblock {\em Quantum}, 5, 11 2021.

\bibitem{yoshioka_generalized_2021}
Nobuyuki Yoshioka, Hideaki Hakoshima, Yuichiro Matsuzaki, Yuuki Tokunaga,
  Yasunari Suzuki, and Suguru Endo.
\newblock Generalized quantum subspace expansion.
\newblock {\em arXiv:2107.02611 [quant-ph]}, July 2021.
\newblock arXiv: 2107.02611.

\bibitem{lowe_unified_2020}
Angus Lowe, Max~Hunter Gordon, Piotr Czarnik, Andrew Arrasmith, Patrick~J.
  Coles, and Lukasz Cincio.
\newblock Unified approach to data-driven quantum error mitigation.
\newblock {\em arXiv:2011.01157 [quant-ph]}, November 2020.
\newblock arXiv: 2011.01157.

\bibitem{aaronson_shadow_2018}
Scott Aaronson.
\newblock Shadow {Tomography} of {Quantum} {States}.
\newblock {\em arXiv:1711.01053 [quant-ph]}, November 2018.
\newblock arXiv: 1711.01053.

\bibitem{huang_predicting_2020}
Hsin-Yuan Huang, Richard Kueng, and John Preskill.
\newblock Predicting many properties of a quantum system from very few
  measurements.
\newblock {\em Nature Physics}, 16(10):1050--1057, October 2020.
\newblock Bandiera\_abtest: a Cg\_type: Nature Research Journals Number: 10
  Primary\_atype: Research Publisher: Nature Publishing Group Subject\_term:
  Information theory and computation;Mathematics and computing;Quantum
  information;Quantum physics;Theoretical physics Subject\_term\_id:
  information-theory-and-computation;mathematics-and-computing;quantum-information;quantum-physics;theoretical-physics.

\bibitem{PhysRevResearch.4.013054}
Hong-Ye Hu and Yi-Zhuang You.
\newblock Hamiltonian-driven shadow tomography of quantum states.
\newblock {\em Phys. Rev. Research}, 4:013054, Jan 2022.

\bibitem{hu_classical_2021}
Hong-Ye Hu, Soonwon Choi, and Yi-Zhuang You.
\newblock Classical {Shadow} {Tomography} with {Locally} {Scrambled} {Quantum}
  {Dynamics}.
\newblock {\em arXiv:2107.04817 [cond-mat, physics:quant-ph]}, July 2021.
\newblock arXiv: 2107.04817.

\bibitem{Ohliger_2013}
M~Ohliger, V~Nesme, and J~Eisert.
\newblock Efficient and feasible state tomography of quantum many-body systems.
\newblock {\em New Journal of Physics}, 15(1):015024, jan 2013.

\bibitem{cotler_quantum_2019}
Jordan Cotler, Soonwon Choi, Alexander Lukin, Hrant Gharibyan, Tarun Grover,
  M.~Eric Tai, Matthew Rispoli, Robert Schittko, Philipp~M. Preiss, Adam~M.
  Kaufman, Markus Greiner, Hannes Pichler, and Patrick Hayden.
\newblock Quantum {Virtual} {Cooling}.
\newblock {\em Physical Review X}, 9(3):031013, July 2019.
\newblock arXiv: 1812.02175.

\bibitem{brandao_local_2016}
Fernando G. S.~L. Brandao, Aram~W. Harrow, and Michal Horodecki.
\newblock Local random quantum circuits are approximate polynomial-designs.
\newblock {\em Communications in Mathematical Physics}, 346(2):397--434,
  September 2016.
\newblock arXiv: 1208.0692.

\bibitem{aaronson_improved_2004}
Scott Aaronson and Daniel Gottesman.
\newblock Improved {Simulation} of {Stabilizer} {Circuits}.
\newblock {\em Physical Review A}, 70(5):052328, November 2004.
\newblock arXiv: quant-ph/0406196.

\bibitem{grassl_bounds_nodate}
Markus Grassl.
\newblock Bounds on the minimum distance of linear codes and quantum codes.

\bibitem{robust_shadow2021}
Senrui Chen, Wenjun Yu, Pei Zeng, and Steven~T. Flammia.
\newblock Robust shadow estimation.
\newblock {\em PRX Quantum}, 2(3), Sep 2021.

\bibitem{2020arXiv201111580E}
Dax {Enshan Koh} and Sabee {Grewal}.
\newblock {Classical Shadows with Noise}.
\newblock {\em arXiv e-prints}, page arXiv:2011.11580, November 2020.

\bibitem{2022arXiv220203272B}
Kaifeng {Bu}, Dax {Enshan Koh}, Roy~J. {Garcia}, and Arthur {Jaffe}.
\newblock {Classical shadows with Pauli-invariant unitary ensembles}.
\newblock {\em arXiv e-prints}, page arXiv:2202.03272, February 2022.

\bibitem{PhysRevLett.106.180504}
Easwar Magesan, J.~M. Gambetta, and Joseph Emerson.
\newblock Scalable and robust randomized benchmarking of quantum processes.
\newblock {\em Phys. Rev. Lett.}, 106:180504, May 2011.

\bibitem{PRXQuantum.2.010351}
Jahan Claes, Eleanor Rieffel, and Zhihui Wang.
\newblock Character randomized benchmarking for non-multiplicity-free groups
  with applications to subspace, leakage, and matchgate randomized
  benchmarking.
\newblock {\em PRX Quantum}, 2:010351, Mar 2021.

\bibitem{levy_classical_2021}
Ryan Levy, Di~Luo, and Bryan~K. Clark.
\newblock Classical {Shadows} for {Quantum} {Process} {Tomography} on
  {Near}-term {Quantum} {Computers}.
\newblock {\em arXiv:2110.02965 [cond-mat, physics:physics, physics:quant-ph]},
  October 2021.
\newblock arXiv: 2110.02965.

\bibitem{2022arXiv220108194S}
Stefan~H. {Sack}, Raimel~A. {Medina}, Alexios~A. {Michailidis}, Richard
  {Kueng}, and Maksym {Serbyn}.
\newblock {Avoiding barren plateaus using classical shadows}.
\newblock {\em arXiv e-prints}, page arXiv:2201.08194, January 2022.

\bibitem{AlirezaQEM}
Alireza Seif, Ze-Pei Cian, Sisi Zhou, Senrui Chen, and Liang Jiang.
\newblock Shadow distillation: Quantum error mitigation with classical shadows
  for near-term quantum processors.
\newblock (To appear).

\end{thebibliography}


\appendix
\onecolumngrid

\section{Stabilizer algorithms}

\subsection{Evaluating the trace in \eqnref{eq:numerator}\label{ap:trace}}

In this appendix, we explain an efficient approach to evaluate the trace in \eqnref{eq:numerator}. For random Clifford ensemble, the reconstruction map reads $\hat{\rho}=\scM^{-1}(\hat{\sigma})=(2^n+1)\hat{\sigma}-\id$, such that
\eq{\dsE_{\hat{\rho}}\Tr\bigg(\Pi \Big(\prod_{s=1}^{m}\hat{\rho}_s\Big) \Pi^\dagger O\bigg)=\sum_{q=0}^{m} \binom{m}{q} (2^n+1)^q(-1)^{m-q}\dsE_{\hat{\sigma}}\Tr\bigg(\Pi \Big(\prod_{s=1}^{q}\hat{\sigma}_s\Big) \Pi^\dagger O\bigg).\label{eq:trace_lst}}
Notice that the projection operator $\Pi=\prod_{j=1}^{n-k}(\id+S_j)/2$ and the snapshot state $\hat{\sigma}=U^\dagger\ket{b}\bra{b}U=\prod_{i=1}^{n}(\id+b_i U^\dagger Z_i U)/2$ both take the from of stabilizer states. So the problem boils down to evaluating the trace of the following general form
\eq{\label{eq:general_Tr}\Tr\bigg(\prod_{j=1}^{l}(a_j\id+b_j M_j)\bigg),}
where $M_j$ are Pauli operators and $a_j,b_j$ are real coefficients. As we expand the product, the only terms that survive the trace are those terms with the Pauli operators multiplied to the identity operator. To find these combination of Pauli operators, we can first encode every Pauli operator $M_j$ as a binary vector following
\eq{M_j=\ii^{\sum_{i=1}^{n}\xi_{ij}\zeta_{ij}}\prod_{i=1}^{n}X_i^{\xi_{ij}}\prod_{i=1}^{n}Z_i^{\zeta_{ij}}\to
\left(\begin{array}{c}
\vdots\\
\xi_{ij}\\
\zeta_{ij}\\
\vdots
\end{array}\right),\label{eq:binary_repr}}
where $\xi_{ij},\zeta_{ij}\in\{0,1\}$ are binary variables. Arranging all the binary vector representations of $M_j$ as column vectors, together they form a $2n\times l$ matrix, denoted as $A$. Each combination of Pauli operators $M_j$ that multiply to identity corresponds to a binary null vector solution $x$ of the binary matrix $A$, as $A x = 0$ (modulo 2). The null vectors form the null space of $A$, denoted as $\scN_A$. The null space of binary matrix $A$ can be found using Gaussian elimination method, and its time complexity is $\scO(nl\times\text{min}(2n,l))$. For $x\in\scN_A$,
\eq{\prod_{j=1}^{l}(M_{j})^{x_{j}}=z(x)\id,}
which defines the phase factor $z(x)$ given $x$. Then the trace in \eqnref{eq:general_Tr} is given by
\eq{\Tr\bigg(\prod_{j=1}^{l}(a_j\id+b_j M_j)\bigg)=2^n\sum_{x\in\scN_A}z(x)\prod_{j=1}^{l}a_j^{1-x_j}b_j^{x_j}.\label{eq:general_trace}} Therefore the time complexity of evaluating the general trace \eqnref{eq:general_trace} is $\scO(nl\times\text{min}(2n,l)+|\scN_A|)$, where $|\scN_A|$ is the volume of the null space $\scN_A$ that is determined by the set of Pauli operators $\{M_i\}$. Applying this result for \eqnref{eq:trace_lst}, we get the time complexity for evaluating $\Tr\left(\Pi\left(\prod^{m}_{s=1}\hat{\rho}_s\right)\Pi^{\dagger}O\right)$ is upper bounded by $\scO(mnl\times\text{min}(2n,l)+m|\scN_A|)$, with $l=mn+n-k+1$. For large $m$, the volume of null space $|\scN_A|$ can be troublesome, and scale exponentially with $m$. But luckily for $m=1$, there exists more efficient polynomial time algorithm, which is illustrated in \appref{ap:stab_projection}.

\subsection{Efficient projection of a stabilizer state\label{ap:stab_projection}}

As shown in \eqnref{eq:binary_repr}, any Pauli string operator can be represented as a one-hot binary vector $x$ and $z$, with $x_i,z_i=0,1$ for $i=1,\dots,N$, where $N$ is the total number of qubits,
\eqs{
\sigma_{(x,z)}=i^{x\cdot z}\prod^{N}_{i=1}X_{i}^{x_i}Z_{i}^{y_i},
}
where $X_i,Z_i$ are Pauli operators, and $x_i, z_i$ are binary values. The multiplication of two Pauli operators can be represented as
\eqs{
\sigma_{(x,z)}\sigma_{(x',z')}=i^{p(x,z;x',z')}\sigma_{(x+x',z+z')\%2},
}
where the phase factor is
\eqs{
p(x,z;x',z')=\sum_{i=1}^{N}\left(z_ix'_i-x_iz'_i + 2(z_i+z'_i)\left\lfloor\frac{x_i+x'_i}{2}\right\rfloor+2(x_i+x'_i)\left\lfloor\frac{z_i+z'_i}{2}\right\rfloor\right)\mod 4.
}
Any two Pauli strings either commute or anti-commute, 
\eq{
\sigma_{(x,z)}\sigma_{(x',z')}=(-)^{c(x,z;x',z')}\sigma_{(x',z')}\sigma_{(x,z)},
}
where the anticommutation indicator $c$ has a simpler form
\eq{
c(x,z;x',z')=\frac{p(x,z;x',z')-p(x',z';x,z)}{2}=\sum_{i=1}^{N}\left(z_ix'_i-x_iz'_i\right)\mod 2.
}
Therefore, the complexity of calculating anticommutation indicator is $\scO(N)$. The binary vectors $x$ and $z$ can be interweaved into a $2N$-component vector $g=(x_0,z_0,x_1,z_1,\cdots)$, which forms the binary representation of a Pauli operator $\sigma_g$.

\begin{figure}[htbp]
    \centering
    \includegraphics[width = 0.4\linewidth]{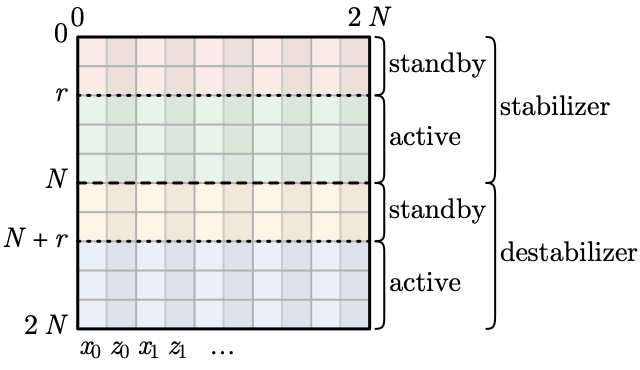}
    \caption{Data structure of a stabilizer state. Each Pauli string is represented as a binary vector. First $N$ rows store the stabilizers of the state, and second $N$ rows store the destabilizers of the state.}
    \label{fig:data_structure}
\end{figure}
In \figref{fig:data_structure}, each row is a binary representation of a Pauli string. For a Hilbert space with dimension $\dsC^{2^N}$, we can find at most $N$ stabilizers $\{S_i,i=1\dots N\}$ and $N$ destabilizers $\{D_i,i=1\dots N\}$, with $[S_i,S_j]=0$, $[D_i,D_j]=0$ and $\{S_i,D_j\}=\delta_{i,j}$. \figref{fig:data_structure} is called the stabilizer tableau of stabilizer state. We use $r$ in \figref{fig:data_structure} indicates the $\log-2$ rank of the density matrix. For pure stabilizer state, $r=0$. And if $r>0$, then stabilizer tableau represents a mixed state and only partial Hilbert space is stabilized. 

A stabilizer projector $\Pi = \prod_{k=1}^{l}\frac{\id\pm G_k}{2}$ can also be represented as a $l$-row tableau (only stabilizer). Now we are going to present an efficient algorithm to calculate $\Tr(\Pi \rho \Pi)$, and update the stabilizer tableau of a full rank stabilizer state $\rho$ according to projection $\Pi$. 

$\bullet$ \textbf{Outline of the algorithm:} First, we set $\text{trace}=1$. Then we scan over every observable $G_k$ in the generator of the operator. For each $G_k$, we continue to scan over all operators in the stabilizer tableau. If the observable $G_k$ anticommute with:
\begin{enumerate}
    \item  At least one active stabilizer (the first of them being $S_p$) $\to$ $G_k$ is an \emph{error} operator that take the state out of the code subspace $\to$ stabilizer tableau need to be updated according to $\id\pm G_k$. And trace will be multiplied by 1/2.
    \item Otherwise, $G_k$ is in the stabilizer group generated by $\{S_k\}$. If the phase factor is compatible, then stabilizer state is eigenstate of $(\id\pm G_k)/2$ with eigenvalue 1, i.e. $(\id\pm G_k)\ket{\psi}/2=\ket{\psi}$; if the phase factor is incompatible, it means $(\id\pm G_k)\ket{\psi}/2=0$.
\end{enumerate}
We see this algorithm has a double loop of $\scO(N)$ items, and each anticommutation check takes $\scO(N)$ time. Therefore, the time complexity of this algorithm is $\scO(N^3)$, where $N$ is the total number of qubits in the system.

\section{Error mitigation capability\label{ap:proof_decom}}
In \eqnref{eq:new_spectral_decom}, we show the density matrix subjected to depolarizing noise is naturally the spectral decomposition form. Here, we provide detailed proof of it. Let $\ket{\psi_0}$ be the ideal quantum state encoded with $[[n,k,d]]$ error correction code. And let $\{\ket{\bar{i}},i=1,\dots,2^k\}$ be the orthonormal basis for the logical space. In general, $\ket{\psi_0}=\sum_{i=1}^{2^k}c_i\ket{\bar{i}}$. We assume the simple depolarizing error for each physical qubit. Then
\eqs{
\rho_\epsilon=(1-p)^N\ket{\psi_0}\bra{\psi_0}+p\rho_1+p^2\rho_2+\cdots,
}
where $\rho_i$ is the density matrix with $i$ local error happened, and $N$ is the system size. For example, 
\eqs{
\rho_1=&(XII\cdots I)\ket{\psi_0}\bra{\psi_0}(XII\cdots I)+(YII\cdots I)\ket{\psi_0}\bra{\psi_0}(YII\cdots I)\\ &+(ZII\cdots I)\ket{\psi_0}\bra{\psi_0}(ZII\cdots I)+
(IXI\cdots I)\ket{\psi_0}\bra{\psi_0}(IXI\cdots I)\\ & +\cdots+ (II\cdots IZ)\ket{\psi_0}\bra{\psi_0}(II\cdots IZ),
}
and 
\eqs{
\rho_2=&(XXI\cdots I)\ket{\psi_0}\bra{\psi_0}(XXI\cdots I)+(XYI\cdots I)\ket{\psi_0}\bra{\psi_0}(XYI\cdots I)\\
&+\cdots +(I\cdots ZZ)\ket{\psi_0}\bra{\psi_0}(I\cdots ZZ).
}
We define the support of a Pauli string as number of Pauli operators that is not the identity operator. Let $P_l$ be a Pauli string operator with non-trivial support $l$. For example, the support of Pauli string $XIZYI$ is three. Then any term in $\rho_l$ can be written as $P_l\ket{\psi_0}\bra{\psi_0}P_l$ with some $P_l$. It is easy to check $\Pi P_l\ket{\psi_0}\bra{\psi_0} P_l \Pi=0$ for any $l<d$. By definition of the code distance $d$, any $P_l$ with $l<d$ is not in the stabilizer group. Therefore, it must anti-commute with some stabilizer generator $S$, such that $\{S,P_l\}=0$. We write $\Pi=\Pi^{'}(\id+S)$, where $\Pi^{'}$ includes all other stabilizer generators. Then
\eq{
\Pi P_l\ket{\psi_0}=\Pi^{'}(\id+S) P_l\ket{\psi_0}=\Pi^{'}P_l(\id-S) \ket{\psi_0}=\Pi^{'}P_l(\ket{\psi_0}-\ket{\psi_0}) =0,
}
and 
\eq{
\Pi P_l\ket{\psi_0}\bra{\psi_0} P_l \Pi=0~~(l<d).\label{eq:lemma1}
}
Therefore, we conclude $\Pi \rho_l \Pi=0$ for any $l<d$, and 
\eqs{
&\dfrac{\Tr(\Pi\rho_{\scE}\Pi O)}{\Tr(\Pi\rho_{\scE}\Pi)}\\ &=\bra{\psi_0}O\ket{\psi_0}\left[1+O\left(p^d\dfrac{\Tr(\rho_d O)}{\bra{\psi_0}O\ket{\psi_0}}\right)\right].
}

Now we want to prove the leading order correction of $\Pi \rho_{\epsilon}^{2}\Pi$ is of order $\scO(p^{2d})$ with contradiction. Suppose the leading order correction is of order $\scO(p^{s})$ with $s<2d$, then there exist $P_l$ and $P_r$ with $l+r=s<2d$ such that 
\eq{
\Pi (P_l\ket{\psi_0}\bra{\psi_0} P_l) (P_r \ket{\psi_0}\bra{\psi_0} P_r) \Pi=\bra{\psi_0} P_l P_r \ket{\psi_0}(\Pi P_l\ket{\psi_0})(\bra{\psi_0} P_r \Pi)\neq 0.
}
This requires $\Pi P_l\ket{\psi_0}\neq 0$ and $\Pi P_r\ket{\psi_0}\neq 0$. From \eqnref{eq:lemma1}, we know this requires $l\geq d$ and $r\geq d$, and it contradicts with $l+r<2d$. Therefore, we conclude the leading order correction of $\Tr(\Pi \rho^{2}_{\epsilon}\Pi O)$ is of order $\scO(p^{2d})$.

For higher power of $\Pi \rho_{\epsilon}^m \Pi$, one may expect the leading order correction will be $\scO(p^{md})$. Depending on the particular logical state $\ket{\psi_0}$ and error correction code, the performance may or may not reach $\scO(p^{md})$. This is because there can exist shortcuts that make the leading order correction larger than $\scO(p^{md})$. In practice, we do witness the performance will be improved with larger $m$.

\begin{figure}[htbp]
    \centering
    \includegraphics[width = 0.35\linewidth]{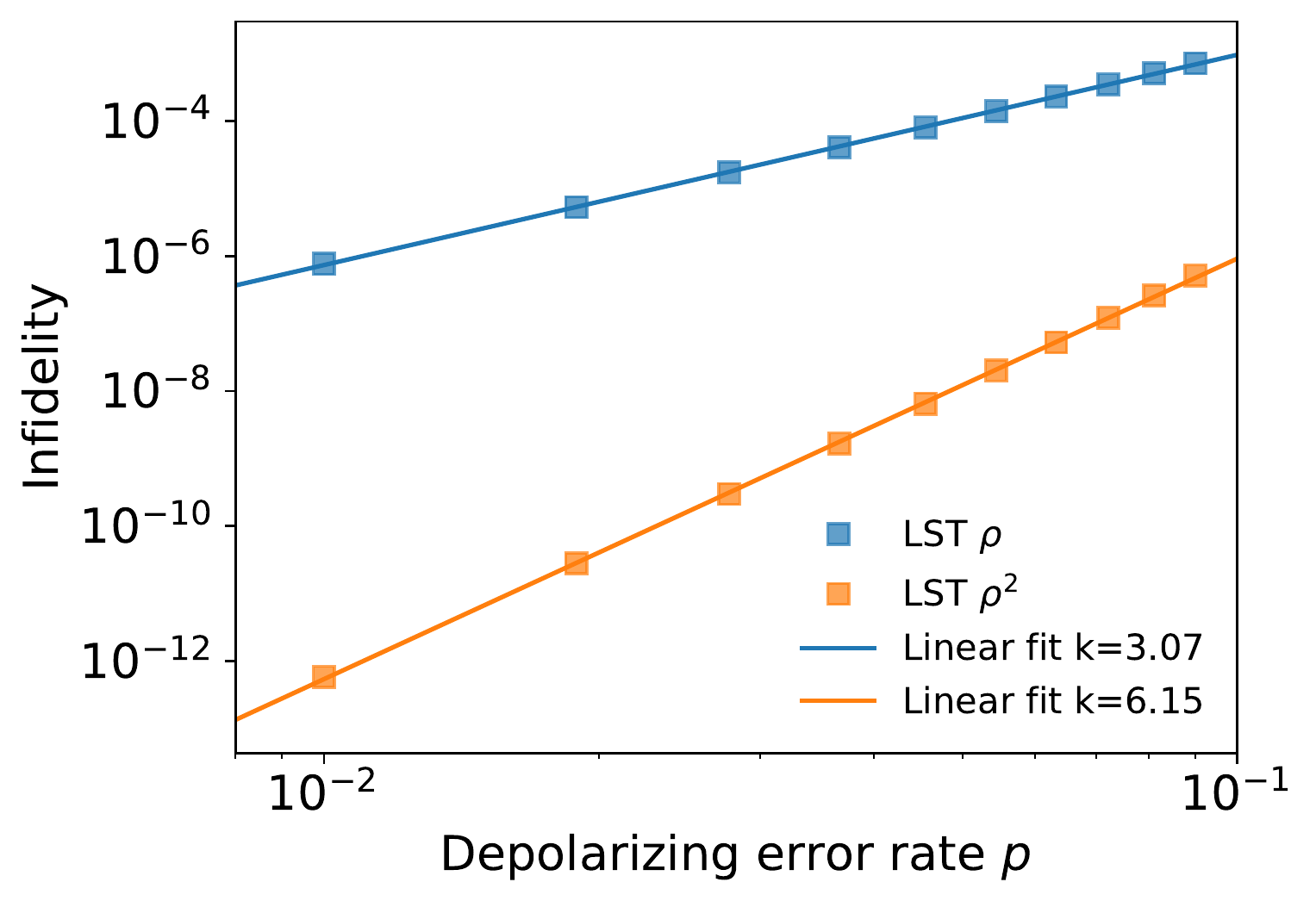}
    \caption{Infidelity in small error rate region. Theoretically we have shown the leading order correction to infidelity will be $\scO(p^{md})$ with $m=1,2$. Here, we use $[[5,1,3]]$ code with LST as a demonstration. We prepare random logical states and calculate the infidelity. We see the numerical results give linear order correction $\scO(p^{3.07})$ and $\scO(p^{6,15})$, which is very close to theoretical prediction $\scO(p^3)$ and $\scO(p^6)$.}
    \label{fig:small_p_expansion}
\end{figure}

\section{Mean and variance of a ratio of two random variables}

Consider random variables $P$ and $Q$ and let $G=g(P,Q)=P/Q$. In general, there is no closed form expression for $\dsE[G(P,Q)]$, and $\var[G(P,Q)]$. Here, we find approximations for the mean and variance using Taylor expansions of $g(P,Q)$.

The approximation for the mean value is 
\eqs{
\dsE[g(P,Q)]&=\dsE[g(\mu_{P},\mu_{Q})+g'_{P}(\mu_{P},\mu_{Q})(P-\mu_P)+g'_{Q}(\mu_{P},\mu_{Q})(Q-\mu_{Q})+R]\\
&\approx \dsE[g(\mu_{P},\mu_{Q})]+g'_{P}(\mu_{P},\mu_{Q})\dsE[(P-\mu_P)]+g'_{Q}(\mu_{P},\mu_{Q})\dsE[(Q-\mu_{Q})]\\
&=g(\mu_{P},\mu_{Q}),
}
where $R$ is the higher order reminders of the Taylor expansion. For keeping the Taylor expansion to the first order, we ignore higher order remainders. 

For the variance, we have
\eqs{
\var[g(P,Q)]&=\dsE\left\{[g(P,Q)-\dsE(g(P,Q))]^2\right\}\\
&\approx \dsE\left\{[g(P,Q)-g(\mu_{P},\mu_{Q})]^2\right\}\\
&\approx \dsE\left\{ [g'_{P}(\mu_P,\mu_Q)(P-\mu_P)+g'_{Q}(\mu_P,\mu_Q)(Q-\mu_Q)]^2\right\}\\
&=g'^{2}_{P}(\mu_P,\mu_Q)\var(P)+g'^{2}_{Q}(\mu_P,\mu_Q)\var(Q)+2g'_{P}(\mu_P,\mu_Q)g'_{Q}(\mu_P,\mu_Q)\cov(P,Q).
}
For our case, $g(P,Q)=P/Q$, therefore $g'_{P}(\mu_P,\mu_Q)=1/\mu_Q$, $g'_{Q}(\mu_P,\mu_Q)=-\mu_{P}/\mu^2_Q$, and 
\eqs{
\var(P/Q)\approx (\dfrac{\mu_P}{\mu_Q})^2\left[\dfrac{\var(P)}{\mu_P^2}+\dfrac{\var(Q)}{\mu_Q^2}-2\dfrac{\cov(P,Q)}{\mu_P\mu_Q}\right]
}

\section{Proof of sample complexities\label{ap:proof}}

Suppose we want to predict a linear property of the underlying quantum state,
\eqs{
o=\Tr(\rho O).
}
We simply replace the unknown quantum state $\rho$ with the classical shadows $\hat{\rho}=\scM^{-1}(\hat{\sigma})$. This yields a stochastic number $\hat{o}=\Tr(\hat{\rho}O)$, and it will converge to correct answer with sufficient amount of classical shadows,
\eq{\mathbb{E}\hat{o}=\Tr(\rho O).}
In practice, the expectation $\mathbb{E}\hat{o}_i$ is replaced by a sample mean estimator, $o_{\avg} = \frac{1}{M}\sum_{i=1}^{M}\hat{o}_i=\frac{1}{M}\sum_{i=1}^{M}\Tr(O\hat{\rho}_i)$. Based on Chebyshev’s inequality, the probability of the
estimation $o_{\avg}$ to deviate from its expectation value $o$ is
bounded by its variance $\var(o_{\avg})$ as $\Pr(|o_{\avg}-o|\geq \delta)\leq \var(o_{\avg})/\delta^2$. To control the deviation within a desired statistial accuracy $\epsilon$, we require $\var(o_{\avg})/\delta^2=\var(\hat{o})/(M\delta^2)\leq \epsilon$, where $M$ is the number of classical shadows. In other words, the number of experiments needed to achieve the statistical error $\epsilon$ is given by
\eq{
M\geq \var(\hat{o})/\epsilon \delta^2.
}
Therefore, the sample complexity is directly related to the variance of single-shot estimation $\var(\hat{o})$. We can further bound the variance by 
\eqs{
\var(\hat{o})&=\dsE[\hat{o}^2]-\dsE[\hat{o}]^2\leq\dsE[\hat{o}^2]\\
&=\dsE_{U\sim\scU}\underset{b\in\{0,1\}^{n}}{\sum}\bra{b}U\sigma U^{\dagger}\ket{b}\bra{b}U\scM^{-1}(O)U^{\dagger}\ket{b}^2\\
&\leq||O||^{2}_{\text{shadow}},
}
where the shadow norm of an observable is defined as 
\eqs{
||O||_{\text{shadow}}&=\max_{\sigma:\text{state}}\left(\dsE_{U\sim\scU}\underset{b\in\{0,1\}^{n}}{\sum}\bra{b}U\sigma U^{\dagger}\ket{b}\bra{b}U\scM^{-1}(O)U^{\dagger}\ket{b}^2\right)^{1/2}\\
&=\max_{\sigma:\text{state}}\left(\dsE_{U\sim\scU}\underset{b\in\{0,1\}^{n}}{\sum}\Tr(\sigma U^{\dagger}\ket{b}\bra{b}U\bra{b}U\scM^{-1}(O)U^{\dagger}\ket{b}^2)\right)^{1/2}\\
&=\max_{\sigma:\text{state}}\left(\Tr\sigma V_{\scU}[O]\right)^{1/2},\label{eq:shadow_norm}
}
where we define a new operator $V_{\scU}[O]=\dsE_{U\sim\scU}\underset{b\in\{0,1\}^{n}}{\sum}U^{\dagger}\ket{b}\bra{b}U\bra{b}U\scM^{-1}(O)U^{\dagger}\ket{b}^2$ that depends both on the unitary ensemble $\scU$ and observable $O$. If the unitary ensemble $\scU$ satisfies unitary 3-design, it can be simplified as 
\eqs{
V_{\scU}[O]&=\dsE_{U\sim\scU}\underset{b\in\{0,1\}^{n}}{\sum}U^{\dagger}\ket{b}\bra{b}U\bra{b}U\scM^{-1}(O)U^{\dagger}\ket{b}^2\\
&=\underset{b\in\{0,1\}^{n}}{\sum}\underset{\sigma,\tau\in S_{3}}{\sum}\text{Wg}[\sigma\tau^{-1}g_0]A[\sigma]B[\tau],\label{eq:Vop}
}
where $\sigma$, $\tau$ are permutations from permutation group $S_3$, $\text{Wg}[g]$ is the Weingarten function of the permutation
group element $g$, $g_0=(2,3)$ is a fixed permutation to match the tensor network connection, and $A[\sigma]$, $B[\tau]$ are
defined as:
\eqs{\dia{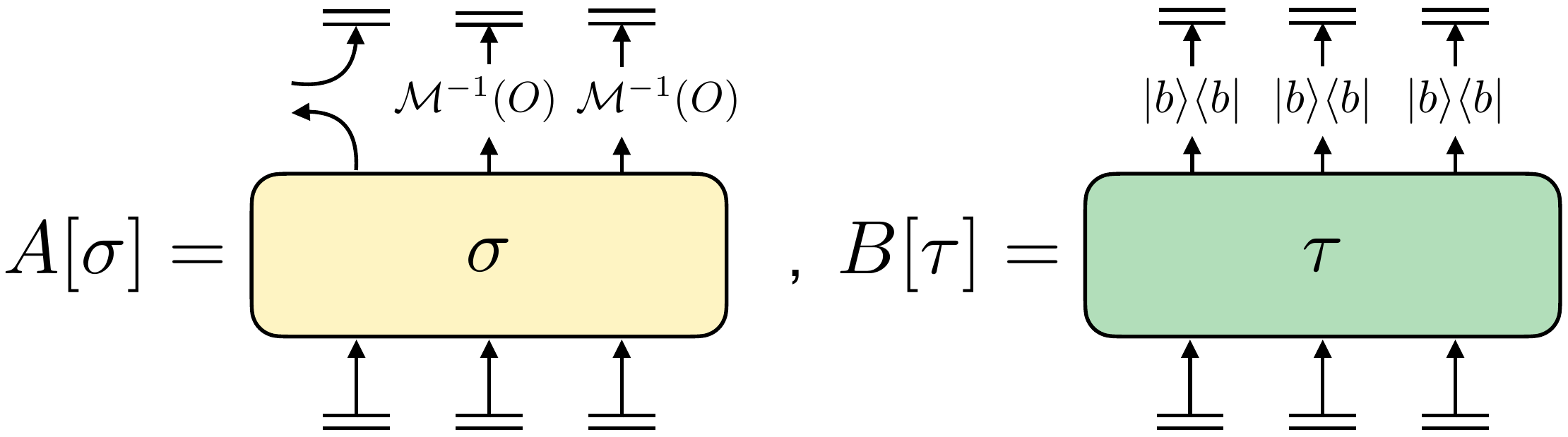}{70}{-35}}
In the following, we will mainly focus on the analysis of $V_{\scU}[O]$ operator and  $||O||^{2}_{\text{shadow}}$. In the main text, we focus on the scheme of encoding each logical qubit with $[n,1]$ stabilizer code, and doing quantum computation with total physical qubits $N=n\times l$, where $l$ is the number of logical qubits. For the classical shadow tomography part, we will use random unitaries sampled from $\Cl(2^n)^{\otimes l}$. One reason of choosing this factorized random unitary group is global clifford group $\Cl(2^{nl})$ is harder to implement in experiments. And the difficulty of implementing this factorized scheme does not depend on number of logical qubits. In practice, it is possible to encode each logical qubit with a small error correction code, such as $[5,1]$ code, and implement random circuits from $\Cl(2^n)$. If the random unitary ensemble is $\Cl(2^n)^{\otimes l}$, then it is easy to show the reconstruction map is
\eqs{
\scM^{-1}[\sigma]=\otimes_{i=1}^{l}\left[(2^n+1)\sigma_i-\Tr(A_i)\id\right],
}
where $\sigma_i$ is the reduced classical shadow on part $i$. The logical Pauli observables will be factorized on each logical sectors, i.e. $O=O_1\otimes O_2\otimes\dots\otimes O_l$. And since the random untaries are sampled from ensemble $\Cl(2^n)^{\otimes l}$, they also have the tensor product structure, i.e. $U=U_1\otimes U_2\otimes\dots \otimes U_l$. By combining those two properties, we can show
\eqs{
V_{\scU}[O]=\otimes_{i=1}^{l}V_{\scU_i}[O_i].
}
Therefore, we only need to focus on the property of $V_{\scU_i}[O_i]$ for each logical sector. In the following, we will use $d=2^n$ for the Hilbert space dimension for one logical sector.

\paragraph{The calculation for $V_{\scU_{i}}[P_i\id_{i}]$}: For projection operator $P_i$, $\scM^{-1}(P_i\id_i)=(d+1)P_i-2\id_i$. And \eqnref{eq:Vop} can be evaluated 
\eqs{
V_{\Cl(d)}[P_i\id_i]=\dfrac{2d-2}{d+2}(P_i+\id_i).
}

\paragraph{The calculation for $V_{\scU_{i}}[P_iO_i]$}: For non-trivial Pauli string $O_i$, $\scM^{-1}(P_i O_i)=(d+1)P_i O_i$. And \eqnref{eq:Vop} can be evaluated 
\eqs{
V_{\Cl(d)}[P_i O_i]=\dfrac{2d+2}{d+2}(P_i+\id_i).
}
As we can see, $V_{\Cl(d)}[P_i\id_i]\lesssim V_{\Cl(d)}[P_i O_i]=\dfrac{2d+2}{d+2}(P_i+\id_i)$. This result indicates the sample complexity for predicting logical Pauli operators $O=\otimes_{i=1}^{l} O_i$ after projection by $P=\otimes_{i=1}^{l} P_i$ does not depend on the locality of the logical Pauli operators, 
\eqs{
||PO||_{\text{shadow}}^2\lessapprox\max_{\sigma:\text{state}}\Tr\left(\sigma \left(\dfrac{2d+2}{d+2}\right)^{l}\otimes_{i=1}^{l}(P_i+\id_i)\right)\lesssim 4^{l}.
}
This result is different from the sample complexity from local Clifford group or tensored Clifford group $\Cl(d)^{\otimes l}$, where the sample complexity will depends on the locality of Pauli string $O$. This difference is mainly introduced by the logical subspace projection $P$. Even the Pauli string $O$ is trivial in some region, the subspace projection $P_i$ will still introduce fluctuation.

\end{document}